# Bayesian estimation of GARCH model by hybrid Monte Carlo

Tetsuya Takaishi[1]

[1]Hiroshima University of Economics, Hiroshima, 731-0192 JAPAN

**Abstract**

The hybrid Monte Carlo (HMC) algorithm is used for Bayesian analysis of the generalized autoregressive conditional heteroscedasticity (GARCH) model. The HMC algorithm is one of Markov chain Monte Carlo (MCMC) algorithms and it updates all parameters at once. We demonstrate that how the HMC reproduces the GARCH parameters correctly. The algorithm is rather general and it can be applied to other models like stochastic volatility models.

**Keywords**: Bayesian inference, GARCH model, Markov chain Monte Carlo, hybrid Monte Carlo, Maximum likelihood estimation.

## 1. Introduction

In order to incorporate the time-vary volatility in financial time series the autoregressive conditional heteroscedasticity (ARCH) model[1] and its generalized version, the GARCH model[2] have been proposed. Usually parameters in the models are inferred by the maximum likelihood estimation or the generalized method of moments [3]. Bayesian inference can be also applied to the GARCH model [4].

Bayesian inference is commonly performed by MCMC algorithms which sample model parameters. The estimates of the model parameters are given by averaging over the sampled parameters.

Popular MCMC algorithms in the Bayesian estimations are the Gibbs sampler ( or heatbath ) and the Metropolis-Hastings algorithm. In updating both algorithms update single parameter each or a block of parameters. Since these algorithms are local updating ones, it is difficult to update all parameters of the model at once.

The HMC algorithm [5] is a global one that can update all parameters at once. Originally the HMC algorithm is proposed for the lattice quantum chromo dynamics (QCD) simulations. The algorithm is, however, not specialized for the lattice QCD simulations but rather general. Here we apply the HMC algorithm to the Bayesian estimation of the GARCH model and demonstrate the HMC estimation of the GARCH parameters.

## 2. Hybrid Monte Carlo

The HMC algorithm [5] consists of molecular dynamics (MD) simulations and Metropolis test. In updating, candidate parameters are given by solving the Hamilton's equations of motion, i.e. MD simulations are performed. During the MD simulations the energy or Hamiltonian $H$ is preserved if the integration is exact. In general the integration is performed approximately which introduces a small energy violation $\Delta H$. This error is corrected at the Metropolis test. If accepted in the Metropolis test, the candidate parameters are kept. Otherwise they are rejected.

Let $f(x)$ be a function proportional to the probability distribution $P(x)$ which we would like to simulate.

$$P(x) = f(x)/Z = \exp(-\ln f(x))/Z,$$

where Z is a normalization constant:

$$Z = \int f(x)dx.$$

With this probability distribution the average value of $g(x)$ is given by

$$\langle g(x) \rangle = \int g(x)f(x)dx/Z$$

Now we introduce $p$ called *momentum* as

$$Z = \int \exp(-p^2/2 + \ln f(x))dxdp$$
$$= \int \exp(-H(x,p))dxdp$$

where $H(x,p) = p^2/2 - \ln f(x)$ is called Hamiltonian. We update $x$ with this joint system of $p$ and $x$. Since the momentum $p$ has no dynamics, the average value of $g(x)$ is unchanged.

The integrator to integrate the Hamilton's equations of motion must satisfy two conditions to maintain detailed balance: (i) simplecticity and (ii) time reversibility. The simplest such integrator is the second order leap frog integrator. Recently a more effective second order integrator for the HMC was found [6]. There is also a possibility to use higher order integrators [7]. In this study we used the simplest one, the second order leap frog integrator.

## 3. GARCH model

The GARCH(m,n) model is given by
$$y_t = \sigma_t \varepsilon_t,$$
$$\sigma_t^2 = \omega + \sum_{i=1}^m \alpha_i y_{t-i}^2 + \sum_{j=1}^n \beta_j \sigma_{t-j}^2,$$
where to ensure positivity of $\sigma_t$, $\alpha_i \geq 0$, $\beta_i \geq 0$, and $\omega > 0$. $\varepsilon_t$ are Gaussian errors with variance=1 and $\langle \varepsilon_t \rangle = 0$.
Here we use GARCH(1,1) model for which $\sigma_t$ is given by
$$\sigma_t^2 = \omega + \alpha y_{t-1}^2 + \beta \sigma_{t-1}^2.$$
For convenience we assume that $y_0$ and $\sigma_0$ are known a priori.

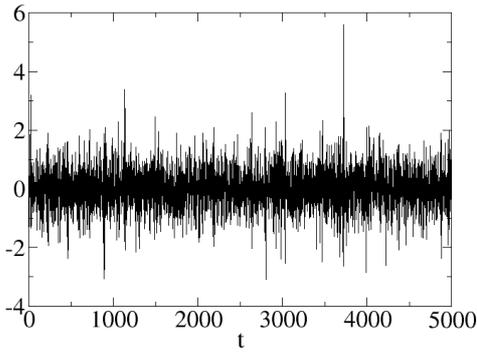

Fig.1: GARCH(1,1) time series with $\alpha$=0.3, $\beta$=0.5 and $\omega$=0.1, generated up to t=5000.

## 4. Bayesian inference

Let $\theta = (\alpha, \beta, \omega)$ be the GARCH parameters to be determined. In the framework of the Bayesian inference the posterior density $\pi(\theta|y)$ with the input data $y$ is given by
$$\pi(\theta|y) \propto f(y|\theta)\pi(\theta)$$
where $f(y|\theta)$ is the likelihood function given by
$$f(y|\theta) = \prod_{t=1}^n \frac{1}{\sqrt{2\pi\sigma_t^2}} \exp\left(-\frac{y_t^2}{\sigma_t^2}\right).$$
We assume that the prior density $\pi(\theta)$ is constant.
In the HMC sampling, we introduce momenta which are conjugate to $\theta = (\alpha, \beta, \omega)$. Then the Hamiltonian is defined as
$$H(\theta, p) = \sum_\theta p_\theta^2/2 - \ln f(y|\theta),$$
and
$$\ln f(y|\theta) = \sum_t \left(\frac{1}{2}\ln \sigma_t^2 - \frac{y_t^2}{\sigma_t^2}\right),$$
where we dropped the irrelevant constant terms.

## 5. HMC Simulations

We generated the GARCH(1,1) time series up to t=5000 with $\alpha$=0.3, $\beta$=0.5 and $\omega$=0.1. Fig.1 shows the time series generated. We prepared 3 sets of data : (1)N=500(t=1 to 500),(2)N=1000(t=1 to 1000) and (3)N=5000(t=1 to 5000), and using these data sets as input we performed HMC simulations to determine the values of $\alpha$, $\beta$ and $\omega$.

|  | true | N=500 | N=1000 | N=5000 |
|---|---|---|---|---|
| $\alpha$ | 0.3 | 0.370(4) | 0.330(2) | 0.2892(3) |
| $\beta$ | 0.5 | 0.460(4) | 0.500(2) | 0.4928(5) |
| $\omega$ | 0.1 | 0.0887(8) | 0.0982(5) | 0.1055(2) |
| $\Delta t$ |  | 0.0002 | 0.0002 | 0.0002 |
| $\tau$ |  | 0.04 | 0.04 | 0.04 |
| Acceptance |  | 0.82 | 0.69 | 0.61 |

Table 1: Summary of results.

The results are summarized in Table 1. $\Delta t$ is the step-size of the leap frog integrator and $\tau$ is the length of the MD trajectory. The acceptance is tuned by the step-size. We do not need to tune the acceptance to a very high value. The optimal acceptance for the second order integrator is about 0.6 to 0.7[7].
The averages of $\alpha$, $\beta$ and $\omega$ are taken over 20000 sampled data. The first 3000 are discarded as thermalization or burn-in period. For N=500, the average values are away from the true values. For N=1000 and 5000, they come close to the true ones.
Fig.2-4 show the time history of sampled $\alpha$ for N=500,1000 and 5000 respectively. The fluctuation reduces as the number of the input data increases.

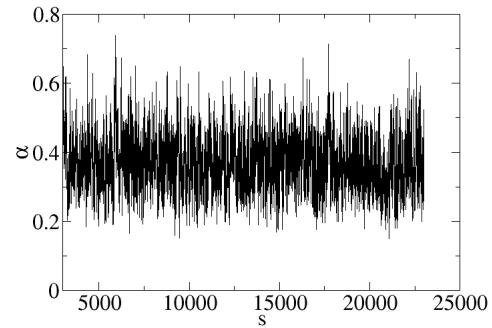

Fig.2: History of sampled $\alpha$ for N=500.

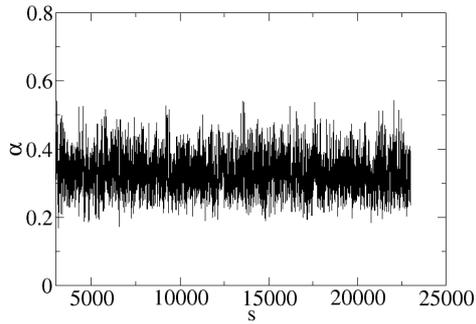

Fig.3: History of $\alpha$ for N=1000.

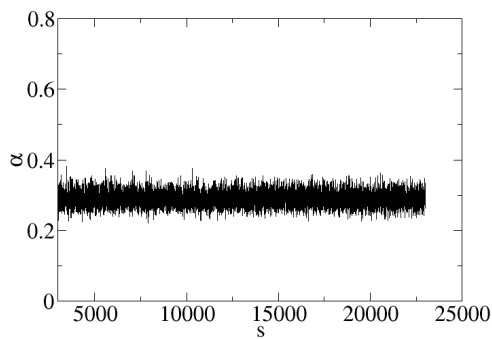

Fig.4: History of $\alpha$ for N=5000.

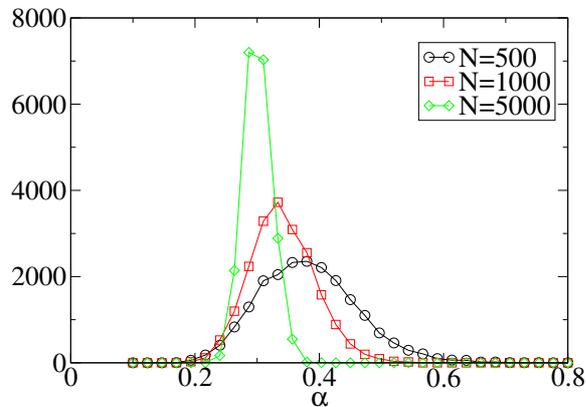

Fig.5: Histogram of sampled $\alpha$.

Fig.5-7 show the histograms of sampled $\alpha$, $\beta$ and $\omega$ respectively. For N=500 the histograms are broad and those peaks are away from the true values. As the number of input data increases the peak moves to the true value.

## 6. Correlations

Fig.8-10 show the scatter plots among the parameters for N=50000. We plotted all data including thermalization. The initial values of the parameters were set to 0.2. We see that starting from 0.2 the parameters move to the thermalized state. There seems to be negative correlation between $\alpha$ and $\beta$, and also $\beta$ and $\omega$. No visible correlation is observed between $\alpha$ and $\omega$.

Fig.11 shows the autocorrelation time of the parameters for N=5000. All the parameters show the similar exponential decay behavior. If we define the correlation time $\tau_0$ by exp(-t/$\tau_0$), we obtain $\tau_0 \approx 3$. Fig.12 shows the autocorrelation time of $\alpha$ for N=500,1000 and 5000. The correlation time seems to increase as the number of input data decreases.

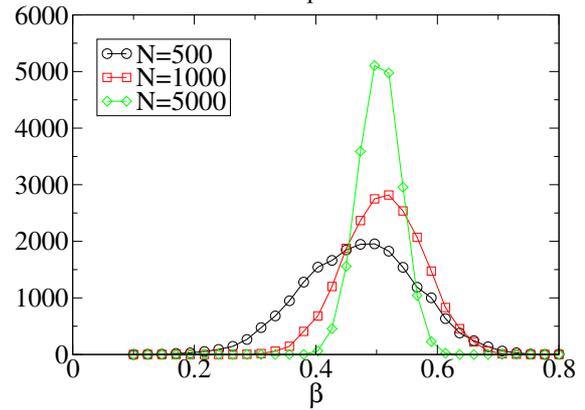

Fig.6: Histogram of sampled $\beta$

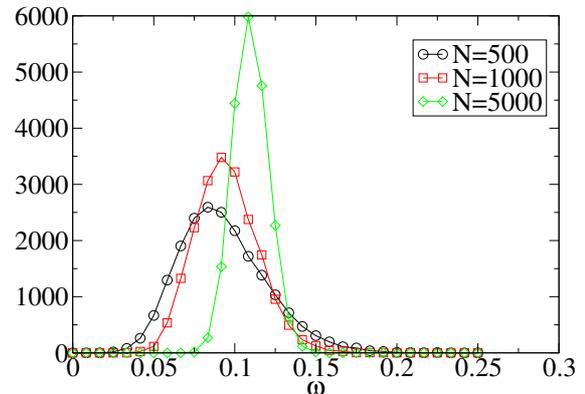

Fig.7: Histogram of sampled $\omega$.

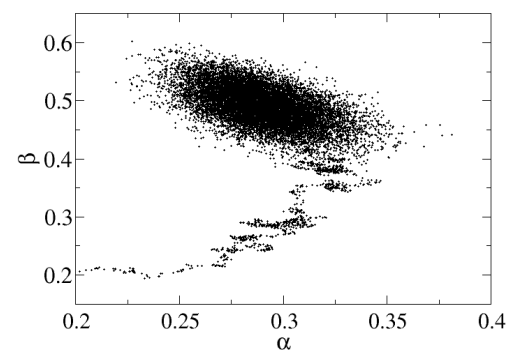

Fig.8: Scatter plot of $\beta$ versus $\alpha$.

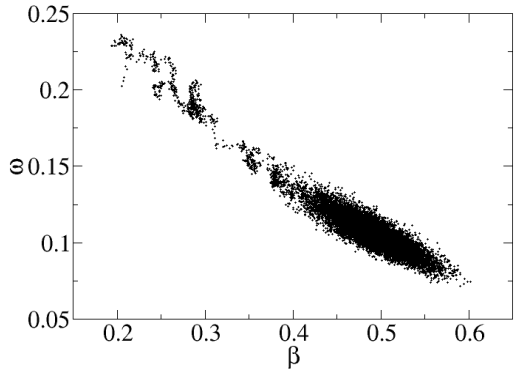

Fig.9: Scatter plot of ω versus β.

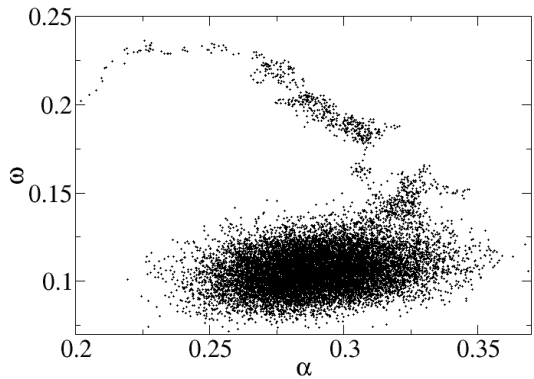

Fig.10: Scatter plot of ω versus α.

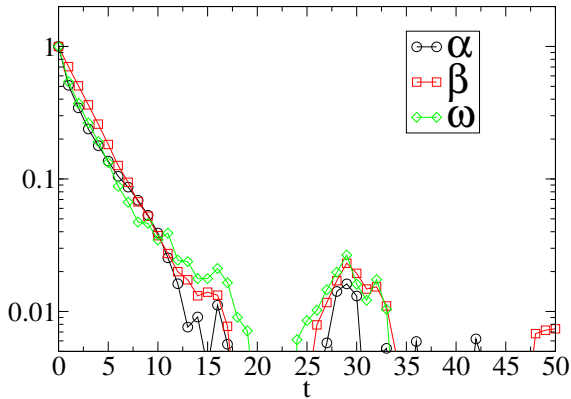

Fig.11: Autocorrelation times of α, β and ω for N=5000.

## 7. Summary

We applied the HMC algorithm to the Bayesian analysis of the GARCH model. In the HMC simulations three parameters of the GARCH model considered here are updated at once. After averaging over the sampled data, we obtain the values close to the true ones. The deviation from the true values decreases for the large number of input data.

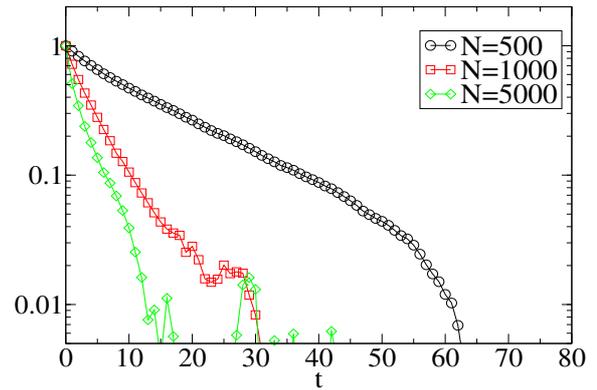

Fig.12: Autocorrelation time of α for N=500,1000 and 5000.

The HMC algorithm is a general method and its implementation is rather simple. It is possible to apply the HMC to other models like stochastic volatility models which have more parameters to be sampled.

The author thanks the Institute of Statistical Mathematics for the use of NEC SX-6.

## 8. References

[1] R.Engel, "Autoregressive conditional heteroscedasticity with estimates of the variance of the UK inflation," Econometrica 50, pp987-1008, 1982.

[2] T.Bollerslev, "Generalized autoregressive conditional heteroscedasticity," J.Econometrics 51, pp.307-327,1986.

[3] See e.g., T.Bollerslev, R.Y.Chou, K.F.Kroner, "ARCH modeling in finance," J.Econometrics 52, pp.5-59, 1992.

[4] L.Bauwens and M.Lubrano, "Bayesian inference on GARCH models using the Gibbs sampler," Econometrics J. 1, pp.23-46.

[5] S. Duane, A.D. Kennedy, B.J. Pendleton, D. Roweth, "Hybrid Monte Carlo," Phys.Lett.B195, pp.216-222, 1987.

[6] T.Takaishi, "Choice of Integrator in the Hybrid Monte Carlo Algorithm," Comput.Phys.Commun. 133, pp.6-17, 2000.
T.Takaishi, "Higher Order Hybrid Monte Carlo at Finite Temperature," Phys.Lett. B540, pp.159-165, 2002.

[7] T.Takaishi and Ph.Forcrand, "Testing and tuning symplectic integrators for Hybrid Monte Carlo algorithm in lattice QCD," Phys. Rev. E 73, pp.036706, 2006.